\documentclass[aps,prb,twocolumn,superscriptaddress,showpacs]{revtex4-1}
\usepackage[utf8x]{inputenc}
\usepackage{amsmath,amssymb,amstext,graphicx,bm}

\usepackage{bbold}
\usepackage[]{natbib}%
\usepackage{hyperref}

\newcommand{\I}{\mathrm{i}}
\newcommand{\E}{\mathrm{e}}

\newcommand{\st}{\mathrm }
\usepackage{color}
\begin{document}
\title{Charge-photon transport statistics and short-time correlations in a single quantum dot-resonator system with arbitrarily large coupling parameter}
\author{T. L. van den Berg}
\affiliation{Centro de F\'isica de Materiales (CFM-MPC) Centro Mixto CSIC-UPV/EHU,
20018 Donostia-San Sebasti\'an, Spain}
\affiliation{Donostia International Physics Center (DIPC), Manuel de Lardizabal 4, 20018 San Sebasti\'an, Spain}

\author{P. Samuelsson}
\affiliation{Department of Physics and NanoLund, Lund University, Box 118, S-221 00 Lund, Sweden}
\date{\today}

\begin{abstract} 
Electrical quantum conductors coupled to microwave resonators have in the last decade emerged as a versatile testbed for controllable light-matter interaction on the nanometer scale. Recent experimental progress with high impedance resonators has resulted in conductor-resonator systems with a large, dimensionless coupling parameter $\lambda \gtrsim 0.1$, well beyond the small coupling regime $\lambda \ll 1$. Motivated by this progress, we here analyse theoretically the joint statistics of transported electrons and emitted photons in a single level quantum dot coupled to a microwave resonator, for arbitrarily large $\lambda$. Describing the electron-photon dynamics via a number-resolved master equation, we evaluate the joint long-time probability distribution as well as joint short-time, $g^{(2)}(t)$, correlation functions. Considering the high-bias regime, with sequential electron tunneling  and working in the damping basis, allows us to obtain analytical results for both transport cumulants and $g^{(2)}(t)$ functions. It is found that the photons emitted out of the resonator are bunched and display a super-Poissonian statistics, for all system parameters. However, the electron transport properties are found to be unaffected by the coupling to the resonator, anti-bunched and with  sub-Poissonian statistics. From the joint distribution we identify regimes of electron tunneling induced photon cascades and very large $g^{(2)}(t)$ functions. All $g^{(2)}(t)$-functions are found to be independent of $\lambda$.  We also identify conditions for and transport signatures of a thermal resonator photon state.
\end{abstract}
\maketitle

\section{Introduction}
Electrons transported through nanoscale conductors, interacting with photons in high-quality microwave resonators, constitute a promising solid state arena for investigations of light-matter interactions in the deep quantum regime. Following early quantum dot-resonator  experiments \cite{Delbecq-2011aa,Frey-2011aa,Petersson-2012aa}, the field has progressed rapidly for almost a decade \cite{Xiang-2013aa,Cottet-2017aa}. Key topics have been e.g., the very recently achieved strong coupling between microwave photons and electron qubits in the charge \cite{Mi-2017ba,Stockklauser-2017aa,Bruhat_2018aa} and spin \cite{Landig-2018aa,Mi-2018aa,Samkharadze-2018aa} degree of freedom, transport induced photon population inversion and lasing (or masing) \cite{Jin-2011aa,Liu-2014aa,Liu-2015aa,Lambert-2015aa,Liu-2017aa,Mantovani-2018aa}, non-classical emitted photon state generation \cite{Leppakangas-2013aa,Gasse-2013aa,Forgues-2014aa,Leppakangas-2015aa,Hassler-2015aa,Grimsmo-2016aa,Westig-2017aa,Rolland-2018aa} and resonator mediated coupling of spatially separated conductors \cite{Childress-2004aa,Bergenfeldt-2013aa,Lambert-2013aa,Contreras-Pulido-2013aa,Xu-2013ba,Delbecq-2013aa,Deng-2015aa,Woerkom_2018aa}. There is also a wide variety of theoretical proposals to use hybrid conductor-resonator systems for e.g., splitting Cooper pairs\cite{Cottet-2012aa, Cottet-2014aa}, detecting Majorana fermions\cite{Trif-2012aa, Schmidt-2013aa,Cottet-2013aa},  and implementing heat engines and refrigerators\cite{Bergenfeldt-2014aa,Hofer-2016aa,Hofer-2016ba}. A majority of the experiments have been carried out with quantum dots in various material systems \cite{Delbecq-2011aa,Delbecq-2013aa,Viennot-2014aa,Liu-2014aa,Liu-2015aa,Liu-2017aa,Viennot-2015aa,Bruhat-2016aa,Frey-2011aa, Frey-2012aa, Toida-2013aa,Basset-2013aa,Stockklauser-2015aa,Mi-2017aa,Petersson-2012aa,Wang-2013aa,Liu-2014aa,Deng-2015aa,Deng-2015ba, Deng-2016aa,Bruhat_2018aa,Mi-2017ba,Stockklauser-2017aa,Landig-2018aa,Mi-2018aa,Samkharadze-2018aa,Woerkom_2018aa}. There are also, including earlier works \cite{Holst-1994aa,Basset-2010aa}, a number of experiments with normal and superconducting tunnel junctions \cite{Hofheinz-2011aa,Gasse-2013aa,Forgues-2014aa,Altimiras-2014aa,Parlavecchio-2015aa,Grimsmo-2016aa,Westig-2017aa,Rolland-2018aa} and an atomic \cite{Janvier-2015aa} point contact.

One important recent development is the application of resonators with large and often tunable characteristic impedance $Z_\text c$, in the k$\Omega$-range, giving rise to a dimensionless electron-photon coupling $\lambda \sim \sqrt{Z_\text c/R_\text Q} \gtrsim 0.1$, where $R_\text Q=25.8$ k$\Omega$ is the quantum resistance. This was used to investigate quantum Fluctuation-Dissipation relations \cite{Parlavecchio-2015aa}, electronic shot noise suppression \cite{Altimiras-2014aa} and, very recently, emitted photon anti-bunching \cite{Rolland-2018aa} in tunnel junction-resonator systems. Moreover, large impedance resonators \cite{Samkharadze-2016aa} were recently used to reach a strong electronic qubit-resonator coupling \cite{Stockklauser-2017aa,Landig-2018aa,Samkharadze-2018aa,Woerkom_2018aa}. Albeit being experimentally challenging, there is no fundamental limitation to reach even larger couplings, $\lambda \sim 1$. As discussed theoretically for a metallic dot \cite{Bergenfeldt-2012aa}, a tunnel junction \cite{Jin-2015aa}, a tunnel-coupled two \cite{Cirio-2016aa} or multi-level \cite{Cirio-2018aa} system and (for moderate coupling strengths) quantum point contact \cite{Mendes-2016aa} conductors,  such large couplings would allow for e.g. electron transport induced, strongly hybridized electron-photon states, i.e., microwave polarons, highly non-equilibrium resonator states, multi-photon creation and destruction, intra photon-mode transitions and ground state electroluminescence. Comparing to e.g., molecular electronic systems, where transport electrons couple strongly to localized phonon modes, conductor-resonator systems thus provide on-chip access to controllable Frank-Condon systems, where the resonator frequency and/or electron tunneling rates, can be tuned in a versatile way. 

Existing experimental techniques allow for an investigation of not only the statistics of electronic transport but also the quantum optics\cite{Walls} properties of the microwave photons emitted from the resonator, such as the long-time average and fluctuations of the photon flow as well as short-time correlation, or $g^{(2)}$, functions. Recent experimental progress towards fast and efficient single microwave photon detection \cite{Chen-2011aa,Inomata-2016aa,Narla-2016aa,Besse-2018aa,Kono-2018aa} also opens up for itinerant microwave photon counting statistics. Various properties of photon emission from hybrid conductor-resonator systems have been investigated both experimentally and theoretically, for a number of conductors such as superconducting \cite{Hofheinz-2011aa,Armour-2013aa, Gramich-2013aa,Trif-2015aa} and normal \cite{Gasse-2013aa,Forgues-2014aa,Xu-2014aa,Jin-2015aa,  Grimsmo-2016aa} tunnel junctions, single \cite{Schiro-2014aa,Dmytruk-2016aa} and double \cite{Xu-2013aa} quantum dots and quantum point contacts \cite{Zakka-Bajjani-2010aa,Hassler-2015aa, Mendez-2016aa,Dambach-2015aa,Leppakangas-2016aa} (see also early works  in \onlinecite{Beenakker-2001aa,Beenakker-2004aa}). 

Taken together, the development towards large electron-photon couplings and fast and efficient detection of both transported electrons and emitted photons, provides a broad scope for investigations of new regimes and novel properties of nano-scale light-matter interactions. In this work we investigate the joint electron-photon transport statistics for a single quantum dot coupled to a microwave resonator with arbitrarily large coupling parameter $\lambda$ (shown schematically in Fig. \ref{fig_sys}). To allow for a largely analytical treatment, we consider the large bias, sequential tunneling regime, where coherences between different photon number states can be neglected. In this regime the system is described by a rate equation, fully accounting for the joint statistical properties of transported electrons and photons. By working in the damping basis, the eigenbasis for spontaneous resonator photon decay, we are able to derive analytical results for the joint long-time (low-frequency) transport cumulants as well as the short-time correlation, $g^{(2)}$- functions, for arbitrary $\lambda$. Moreover, in some limiting cases the full joint probability distribution can be found analytically, within a saddle point approximation. 

Summarizing our key findings, based on the long time statistics we identify parameter regimes for uncorrelated as well as cascaded emission of photons. Leaving the emitted photons unobserved, the electron transport statistics is the same as in the absence of the resonator. The electron and photon statistics are found to be sub- and super-Poissonian, respectively, for all system parameters. For the short time correlations we find that all $g^{(2)}$-functions are independent of $\lambda$. While the electron-electron correlation function describes anti-bunching, $g^{(2)} \leq 1$ the photon-photon function describes bunching $g^{(2)} \geq 1$, i.e., classical light emission for all system parameters. In the regime with photon cascades the photonic correlation function can become very large,  $g^{(2)}\gg 1$. The electron-photon cross correlations depends on the order of the measurement, i.e., whether electrons or photons are measured first. While the resonator photon state typically is in non-equilibrium, we identify a parameter regime for which this state is a thermal state, with corresponding photon emission properties.

The remainder of the paper is organized as follows: In the next section, \ref{sec_system}, we describe the dot-resonator system, define system parameters and specify the working regime. In section \ref{sec_theory} we present the master equation approach to charge and photon transport statistics as well as the dynamics, for arbitrarily large dot-resonator coupling. We then discuss the results, first the long time statistics in section \ref{sec_longt} and then the short time correlation functions in section \ref{sec_g2}. A comparison to existing results for related systems is performed in section \ref{sec_compare}. We end the paper with a conclusion and discussion, section \ref{sec_conclusion}. Some calculations are detailed in the appendix.

\section{System}
\label{sec_system}

\begin{figure}
\includegraphics[width=0.56\linewidth]{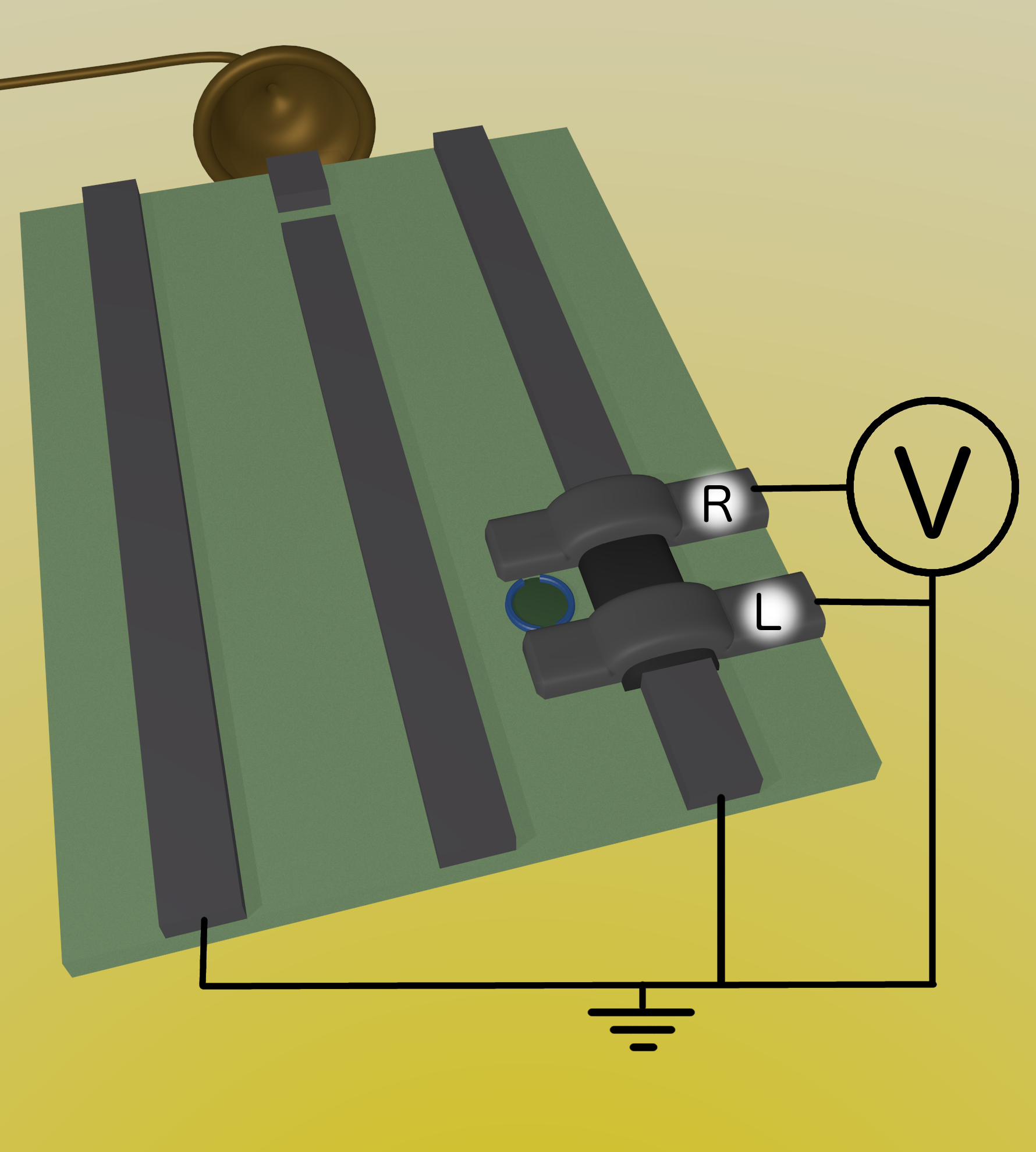}\hskip0.3cm
\includegraphics[width=0.34\linewidth]{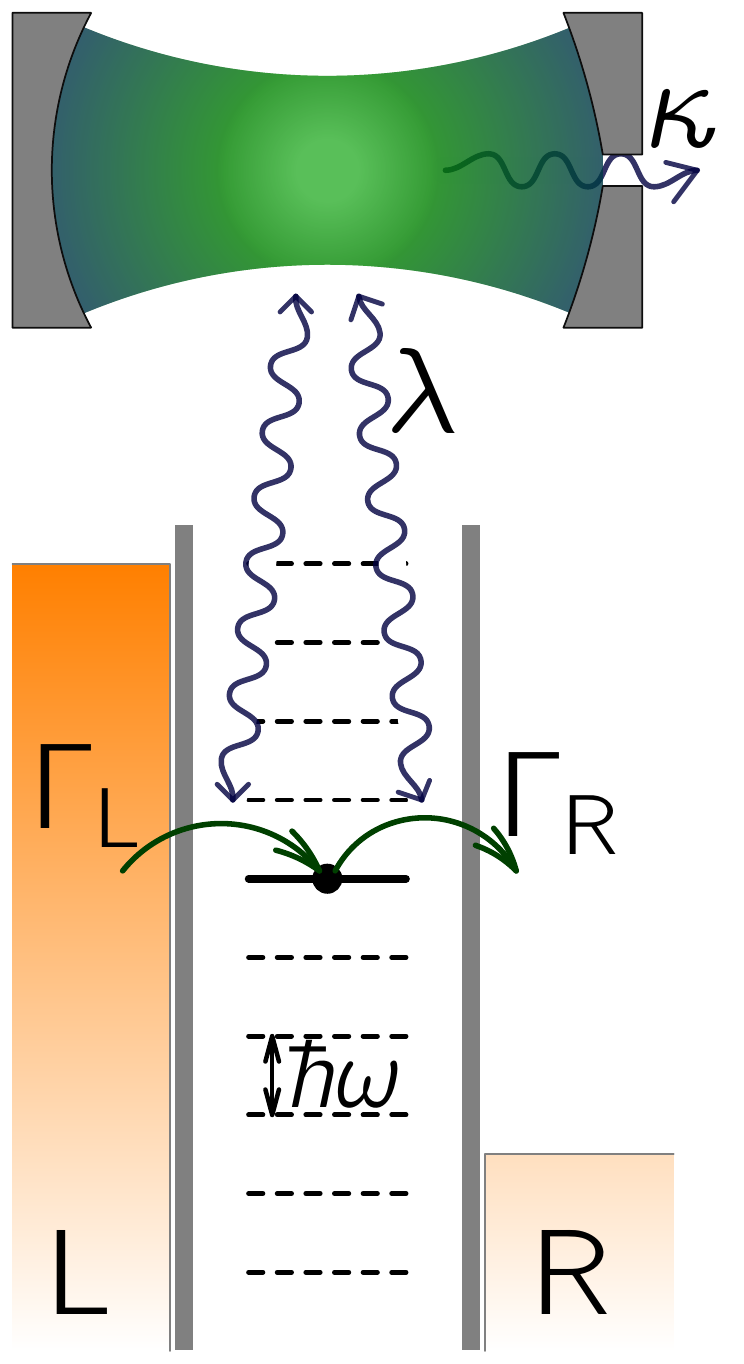}
\caption{\label{fig_sys} Left panel: Schematic of the system. A quantum dot, tunnel coupled to two electronic leads L and R,  is embedded in a coplanar transmission line resonator. A potential bias $V$ is applied symmetrically across the dot. Microwave photons leaking out of at one end of the resonator are collected at a photodetector. Right panel: Electrons tunnel sequentially in and out of the single level dot, with bare rates $\Gamma_\text L$ and $\Gamma_\text R$ respectively. During both the in and the out tunneling the electron can emit or absorb photons of energy $\hbar \omega$ from the resonator, illustrated by introducing a set of effective energy levels  (thin dashed lines) equally spaced around the bare electronic level (thick, solid line). The electron-photon coupling strength is characterized by the dimensionless parameter $\lambda$ and photons in the resonator leak out at a rate $\kappa$.}
\end{figure}

The system under consideration, shown in Fig. \ref{fig_sys}, consists of a quantum dot embedded in a coplanar microwave transmission line resonator. The dot is connected to two electronic leads, L and R,  via barriers characterized by tunnel rates $\Gamma_\text L, \Gamma_\text R$, respectively. A bias $V$ is applied across the dot. We consider a single active level in the dot, with an energy in the middle of the bias window. The charge on the dot is coupled capacitively to a single resonator mode, with frequency $\omega$. The electron-photon coupling rate is $\lambda \omega$, where $\lambda$ is the dimensionless coupling strength. Importantly, here we make no assumptions on the magnitude of $\lambda$, arbitrarily large coupling strength is allowed.
Throughout the paper we work in the high bias regime, $eV \gg \hbar \omega$, as illustrated in Fig.~\ref{fig_sys}. Electrons tunneling through the dot can thus emit or absorb zero, one or multiple photons in both the in-, and the out-tunneling processes, without energy constraints. However, the corresponding tunnel rates are renormalized by the coupling strength $\lambda$, further discussed below.  

The temperature $T$ of the system, including the electronic leads, is taken to fulfill the condition $\hbar  \Gamma_\st{L},\hbar \Gamma_\st{R} \ll k_\text B T \ll \hbar \omega$. Consequently, the electron tunneling is in the sequential regime, {\it i.e.}, cotunneling can be neglected. In addition, we can ignore the coherent dynamics between states with different numbers of resonator photons and describe the transport properties within a rate equation formalism. We account for leakage of photons out of the resonator at a rate $\kappa$. As a consequence of the condition $k_\text B T \ll \hbar \omega$, no photons are injected back into the resonator. We moreover neglect internal resonator losses, supposed to have rates much smaller that \(\kappa\). 

Photons emitted from the resonator are counted in an ideal photocounter shown schematically in Fig. \ref{fig_sys}, the effect of non-unity detector efficiency is discussed further below. The time-averaged electron current and the current fluctuations \cite{Blanter-2000aa} can be measured via the electrical currents at the leads L and R. To experimentally access the full statistics as  well as the short-time correlators of the electron current, real-time detection of the charge in the quantum dot is required \cite{Gustavsson-2006aa} (charge detector not shown in the figure).  

\section{Theory and method}
\label{sec_theory}
Under the conditions stated above, the joint electron-photon dynamics can be described by a rate equation \cite{Mitra-2004aa,Bergenfeldt-2012aa} for the probabilities \(P^n_p(n_\st e , n_\st p , t) \), to have $n=0$ or $1$ electrons on the dot and $p=0,1,2,....$ photons in the resonator, conditioned on that \(n_\st e \) electrons have been transferred through the dot and \(n_\st p \) photons have been emitted out of the resonator at time t. To account for the statistics of \(n_\st e \) and \(n_\st p \), we consider the rate equation in Fourier space \cite{Bagrets-2003aa,Flindt-2005aa,Kiesslich-2006aa} with the variables, or counting fields, \(\xi_\text e\) and \(\xi_\text p\), conjugate to \(n_\st e \) and \(n_\st p \) respectively. In a compact form we can then write
\begin{equation}
\frac{d\bm P(\xi_\text e,\xi_\text p) }{dt} = M(\xi_\text e,\xi_\text p) \bm P (\xi_\text e,\xi_\text p) 
\end{equation}
Here \(\bm P\) is a vector with transformed probabilities, as
\begin{align}
\bm P(\xi_\text e,\xi_\text p) = \begin{pmatrix} \bm P ^0(\xi_\text e,\xi_\text p) \\ \bm P ^1(\xi_\text e,\xi_\text p) \end{pmatrix}, \quad \bm P ^n = \begin{pmatrix} \vdots \\ P^n_p(\xi_\text e,\xi_\text p) \\ P^n_{p+1}(\xi_\text e,\xi_\text p)\\ \vdots \end{pmatrix}\,,
\end{align}
and the evolution matrix can be written
\begin{equation}
\label{eq_Mxichi}
M(\xi_\text e,\xi_\text p)= M_\text e (\xi_\text e) + M_\text p (\xi_\text p)- M_\text 0.
\end{equation} 
The partial electron and photon evolution matrices \(M_\st e \) and \(M_\st p \) can be written as tensor products between matrices in charge- and photon number spaces. For the dot charge matrix we have
\begin{align}
\label{eq_Me}
M_\st e (\xi_\text e) = \mathcal{J}_\st e (\xi_\text e) \otimes M_\lambda\,,
\end{align}
with
\begin{align}
\mathcal{J}_\st e (\xi_\text e) = \begin{pmatrix} 0 & \Gamma_\text R \E^{i \xi_\text e} \\ \Gamma_L  &0 \end{pmatrix}\, 
\end{align}
and
\begin{align}
 M_\lambda = \E^{-\lambda^2} \begin{pmatrix} 1 &  \lambda^2 & \dots \\ \lambda^2  & (1-\lambda^2)^2 &\dots \\ \vdots & \vdots & \ddots \end{pmatrix} \,. \nonumber
\end{align}
The matrix \(M_\lambda\) contains the Franck-Condon factors, that is, the element $[M_\lambda]_{k+1,q+1}$ gives the probability that the number of photons in the resonator changes from $q$ to $k$ during an in- or out- electron tunneling process\cite{Mahan-book, Mitra-2004aa}. The photon evolution matrix can be written 
\begin{align}
M_\st p (\xi_\text p) = \mathcal{J}_\st p (\xi_\text p) \otimes M_\kappa\,,
\end{align}
with
\begin{equation}
 \mathcal{J}_\st p (\xi_\text p) = \E^{i \xi_\text p} \begin{pmatrix} 1 & 0 \\ 0 & 1 \end{pmatrix} \,, \quad
 M_\kappa =  \begin{pmatrix} 0 & \kappa & 0 & 0& \dots \\ 0  & 0 & 2\kappa & 0& \dots \\ 0& 0& 0& 3\kappa & \dots \\ 0& 0& 0& 0 & \dots \\ \vdots & \vdots & \vdots & \vdots & \ddots \end{pmatrix} \,. \nonumber
\label{matrixkappa}
\end{equation}
For $M_\kappa$, the non-zero element $[M_\kappa ]_{q+1,q+2}=(q+1)\kappa$ gives the rate for the number of photons in the resonator to change from $q+1$ to $q$, due to emission of one photon out of the resonator. Lastly, \(M_\st 0 \) is a diagonal matrix guaranteeing probability conservation, such that \(\bm v _0 M (0,0) = 0\) for the row vector \(\bm v _0 =(1,1, \dots )\). 

\subsection{Low frequency statistics}
In the limit of long measurement times  $t$, the total probability \(P(n_\st e,n_\st p )=\sum_{n,p}P^n_p(n_\st e,n_\st p,t)\) can be written as
\begin{align}
\label{eq_Pnpne}
P (n_\st e,n_\st p ) = \frac{1}{(2\pi)^2} \iint_{-\pi}^\pi d \xi_\text e d \xi_\text p \E^{- \I n_\st e \xi_\text e - \I n_\st p \xi_\text p + tF(\xi_\text e, \xi_\text p)}.
\end{align}
Here \(F(\xi_\text e,\xi_\text p)\) is the generating function for the electron and photon low frequency cumulants, obtained from the eigenvalue equation
\begin{align}
\label{eq_MP=FP}
M(\xi_\text e,\,\xi_\text p) \bm P (\xi_\text e,\xi_\text p) = F (\xi_\text e,\xi_\text p) \bm P (\xi_\text e,\xi_\text p) 
\end{align}
picking the eigenvalue going to zero as \(\xi_\text e,\xi_\text p \rightarrow 0\). 

We note that only in limiting cases can the cumulant generating function \(F(\xi_\text e,\xi_\text p)\) be obtained analytically. However, all electron, photon and cross cumulants can be obtained from \(F(\xi_\text e,\xi_\text p)\) by successive derivatives with respect to \(\xi_\text e\) and \(\xi_\text p\). Focusing on the experimentally most accessible, lowest order cumulants, we have the average currents \(I_\st e\) and \(I_\st p \), given by
\begin{align}
I_\st a = - i \sigma_\text a \left. \frac{\partial F(\xi_\text e,\xi_\text p)}{\partial \xi_\text a} \right|_{\xi_\text e=\xi_\text p=0}
\end{align}
with $\text a \in \{\text e,\text p\}$ and $\sigma_\text e=e$ ($e$ is electron charge), $ \sigma_\text p=1$, and the correlations \(S_\st{e,e},\, S_\st{p,p}\) and \(S_\st{e,p}\), given by
\begin{align}
S_\st {a,b} = - \sigma_\text {a,b} \left. \frac{\partial^2 F(\xi_\text e,\xi_\text p)}{\partial \xi_\text a \partial \xi_\text b} \right|_{\xi_\text e=\xi_\text p=0}
\end{align}
with $\text a,\text b \in \{\text e,\text p\}$, $\sigma_\text{e,e}=e^2, \sigma_\text{e,p}= \sigma_\text{p,e}=e$ and $\sigma_\text{p,p}=1$. By expanding Eq. (\ref{eq_MP=FP}) in $\xi_\text e,\xi_\text p$, following  [\onlinecite{Braggio-2009aa},\onlinecite{Flindt-2010aa}], we can express the cumulants in terms of the partial evolution matrices and the steady state probability vector $\bm P=\bm P(0,0)$, formally the right eigenvector to \(M(0,0)\) corresponding to the eigenvalue \(0\). This gives for the current
\begin{align}
\label{eq_IaExpansion}
I_\text a = \sigma_\text a \bm v_0 M^\prime_\text a \bm P , \quad M^\prime_\text a=\left.  \frac{d M_\text a}{d \xi_\text a}\right|_{\xi_\text a=0}.
\end{align}
For the correlators we have
\begin{eqnarray}
\label{eq_SabExpansion}
S_\text {a,b} &=& \sigma_\text{a,b} \bm v_0 \left[M^{\prime\prime}_{\text a,\text b} \bm P + (M^{\prime}_\text a  - I_\text a) \bm P ^{\prime}_\text a+(M^{\prime}_\text b  - I_\text b) \bm P ^{\prime}_\text b \right], \nonumber \\
\bm P^{\prime}_\text a &=& M_\st g ^{-1} (I_\text a - M^{\prime}_\text a) \bm P+ (\mathbb 1 - M_\st g ^{-1} M) \bm w  
\end{eqnarray}
with $M^{\prime\prime}_{\st a,\st b} =\left.  \partial^2 M/ (\partial \xi_\text a \partial \xi_\text b )\right|_{\xi_\text a=\xi_\text b=0}$ (note that $M^{\prime\prime}_{\text e,\text p}=M^{\prime\prime}_{\text p,\text e}=0$), \(M_\st g ^{-1}\) the generalized, Moore-Penrose inverse of the evolution matrix \(M(0,0)\), and  \(\bm{w}\)  any vector that ascertains the relation \(\bm v_0 \bm P^{\prime}_a  =0\).

In some cases it is illustrative to consider the probability distribution function \(P(n_\st e,n_\st p)\) directly. We evaluate the  distribution function in the saddlepoint approximation, to exponential accuracy. This gives the large deviation function
\begin{align}
\label{eq_saddlepoint}
\st{ln}[P(n_\st p , n_\st e )] = t F(\xi^*_\text e,\xi^*_\text p ) - i n_\st e \xi^*_\text e - i n_\st p \xi^*_\text p \,,
\end{align}
where \(\xi^*_\text a=\xi^*_\text a(n_\st e, n_\st p)\) are solutions of the coupled saddle point equations
\begin{align}
\left . \frac{\st \partial F(\xi_\text e, \xi_\text p)}{\st \partial \xi_\text a} \right|_{\xi_\text e= \xi^*_\text e,\xi_\text p =\xi_\text p^*} = \frac{in_\st a}{t}. \label{eq_chisaddle}
\end{align}
In case the generating function \(F(\xi_\text e,\xi_\text p)\) is not analytically available, a systematic, numerically tractable way to solve the saddle point equation is to extend the approach of [\onlinecite{Flindt-2010aa}] to the two dimensional \((\xi_\text e, \xi_\text p)\) space.

\subsection{Short time correlations}
To investigate the short time statistics, we calculate the two point (Glauber) correlation function \cite{Glauber-1963aa,Emary-2012aa}, defined as
\begin{align}
\label{eq_g2definition}
g^{(2)}_{\text{a,b}} (t)= \frac{ \bm v _0 \,  M^\prime_\text a \E^{-M(0,0) t} M^\prime_\text b \, \bm P }{I_\text a I_\text b}\,.
\end{align}
The function \(g_{\text{a,b}}^{(2)}(t)\) gives us information about the probability of measuring a particle \(a\) at time \(t\), given that one has already measured a particle \(b\) at time \(t=0\). We stress that, in our definition of \(g_{\text{a,b}}^{(2)}(t)\) , electrons are measured when they enter the dot and photons are measured when they leave the resonator. Note that, in general, the cross correlation function \(g_{\text{a,b}}^{(2)}(t)\neq g_{\text{b,a}}^{(2)}(t)\), demonstrating that the result depends on the order of measurements. 
 
We note, for completeness, that the short time correlation functions \(g_{\text{a,b}}^{(2)}(t)\) are related to the low frequency correlations \(S_{\text{a,b}}\) by the following relation
\begin{align}
\label{eq_Sabisintg2}
S_{\text{a,b}} = \sigma_\text a I_\text a \delta_{\text a,\text b} + I_\text a I_\text b \int_0^\infty \! dt \, \bigg[ \frac{g^{(2)}_{\text a,\text b}(t)+g^{(2)}_{\text b,\text a}(t)}{2} - 1 \bigg]
\end{align}
where \(\delta_{\text{a,b}}\) is the Kronecker-delta. 

\subsection{Rotating to the damping basis}
\label{subsec_dampingbasis}
As is clear from Eqs. (\ref{eq_IaExpansion}), (\ref{eq_SabExpansion}) and (\ref{eq_g2definition}), the evaluation of the low frequency cumulants and the short time correlations require the knowledge of the steady state probability vector \(\bm P \), obeying the equation \(M(0,0) \bm P =0\). In matrix block form, this equation can be written
\begin{align}
\label{eq_steadystate}
\bigg[ \begin{pmatrix} -\Gamma_\text L \mathbb 1 & \Gamma_\text R M_\lambda \\ \Gamma_\text L M_\lambda & -\Gamma_\text R \mathbb 1 \end{pmatrix} + \begin{pmatrix} \kappa X &0 \\ 0& \kappa X\end{pmatrix} \bigg] \begin{pmatrix} \bm P ^0 \\ \bm P ^1 \end{pmatrix} = 0\,,
\end{align}
where we introduced the two-band diagonal matrix \(X\), with elements \(X_{kk}=-(k-1),\, X_{k,k+1}=k\), and zero otherwise and $\bm P^n=\bm P^n(0,0)$ for $n=0,1$.

Despite the infinite dimensionality of this set of linear equations, it is possible to find closed, analytical expressions for the elements of $\bm P^n$ for arbitrarily large $\lambda$. To arrive at these expressions, it is convenient to transform Eq. (\ref{eq_steadystate}) to the damping basis \cite{Briegel-1993aa, Barnett-2000aa} in photon probability space. This is the basis in which the matrix \(X\) describing the photon leakage out of the resonator is diagonal. The non-orthogonal damping basis transformation is described by the upper triangular matrix \(L\), with elements given in terms of the binomial coefficients as  \([L]_{kq}=(-1)^{k+q} {{q-1}\choose{k-1}}\) for \(q>k\). The relevant matrices and vectors transform as
\begin{align}
X=L^{-1}\bar D L,\quad M_\lambda= L^{-1}\bar M _\lambda L,\quad\bm P ^n = L \bar{\bm P}^n\,,
\label{damptrans}
\end{align}
where the diagonal matrix \(\bar D\) has elements \([\bar D]_{kk}=k-1\), and the lower triangular matrix \([\bar M _\lambda]_{kq}=\lambda^{2(k-q)} {{k-1}\choose{q-1}}\) for \(k>q\). We can then write Eq.~(\ref{eq_steadystate}) as
\begin{align}
\begin{pmatrix} -\Gamma_\text L \mathbb 1 + \kappa \bar D & \Gamma_\text R \bar M _\lambda \\ \Gamma_\text L \bar M _\lambda & -\Gamma_\text R \mathbb 1 + \kappa \bar D \end{pmatrix} \begin{pmatrix} \bar{\bm P} ^0 \\ \bar{\bm P} ^1 \end{pmatrix} = 0\,.
\end{align}
Noting that we can write \(\bar M _\lambda = \sum_k \lambda^{2k} \bar M _\lambda ^{(k)}\), where \(\bar M _\lambda^{(k)}\) has non-zero elements only on the lower \(k\)-th off diagonal, it follows that the transformed probability vector can be written as 
\begin{align}
\label{eq_barP}
\bar{ \bm P} ^n = \sum_{k} \lambda^{2k} c^n_k  \bm e _{k+1} \,,  
\end{align}
with \(\bm e _k \) the canonical basis vector with elements \([\bm e _k ]_q=\delta_{k,q}\), and the coefficients \(c^n_p\) are functions of \(\Gamma_\text{L}, \Gamma_\text{R}\) and \(\kappa\) but independent of \(\lambda\). Explicit expressions for the lowest order coefficients are given in the Appendix.

We note that the expressions for the cumulants, Eqs. (\ref{eq_IaExpansion}) and  (\ref{eq_SabExpansion}), and short time correlators, Eq. (\ref{eq_g2definition}) in the damping basis are found by transforming the matrices in electron-photon space as $M_\text a \rightarrow \bar M _a = ( \mathbb 1 \otimes L^{-1}) M_\text a  ( \mathbb 1 \otimes L)$ etc. and the vector $\bm v_0 \rightarrow  \bar{\bm v }_0 = \bm v _0 (\mathbb 1 \otimes L)=(1,0,0,0.....,1,0,0,0,...)$.

\subsection{Resonator state}
To obtain a better understanding of the transport properties of the system, it is helpful to first analyse the photon state of the cavity, i.e the total probability $P_p=P_p^0+P_p^1$ to have $p$ photons in the resonator. The average number of photons, $\langle N \rangle=\sum_p p P_p$ can be obtained by comparing the rates for photon injection and emission. In the high-bias limit considered, the average number of photons added to the resonator at an electron tunneling event is $\lambda^2$, independent of the state of the resonator. This formally follows from $\sum_{k} k[M_\lambda]_{k+q,q}=\lambda^2$ for all $q$. Moreover, the average rate of electrons tunneling through the dot is $(1/\Gamma_\text L+1/\Gamma_\text R)^{-1}=\Gamma_\text L \Gamma_\text R/(\Gamma_\text L+ \Gamma_\text R)$, independent of the resonator state as further discussed below. The photon injection rate is thus $2\lambda^2\Gamma_\text L \Gamma_\text R/(\Gamma_\text L +\Gamma_\text R)$, with the factor of two accounting for the fact that the electron adds photons both when tunneling in and out of the dot. The emission rate, clear from Eq. (\ref{matrixkappa}), is given by $\kappa \sum_p p P_p=\kappa \langle N \rangle$. In steady state the rates are the same and we have
\begin{equation}
\langle N \rangle=\frac{2\lambda^2\Gamma_\text L \Gamma_\text R}{\kappa(\Gamma_\text L+ \Gamma_\text R)}
\end{equation}
valid for arbitrary $\lambda,\kappa,\Gamma_\text L, \Gamma_\text R$.  We note that for small resonator leakage $\kappa$, the average number of photons $\langle N \rangle \gg 1$, requiring the applied bias to be very high in order to have all photon absorption and emission events during electron tunneling energetically allowed.

In most parameter regimes, the resonator photons are in a non-equilibrium state. However, in the regime $\kappa \ll \Gamma_\text L,\Gamma_\text R$, the resonator state is in thermal equilibrium, characterized by an effective temperature $T_\text{eff}$. This can be shown via  $\bar P_p^0$ and  $\bar P_p^1$ in the damping basis. From the Appendix we have the coefficients in Eq. (\ref{eq_barP}) as $c_{p+1}^0=\langle N \rangle c_{p}^0$ and $c_{p+1}^1=\langle N \rangle c_{p}^1$ for $p \geq 0$. Together with $c_{0}^0=\Gamma_\text R/(\Gamma_\text L+\Gamma_\text R)$ and $c_{0}^1=\Gamma_\text L/(\Gamma_\text L+\Gamma_\text R)$ we get the total $\bar P_p=\bar P_p^0+\bar P_p^1=\langle N \rangle^p$. Transforming back, via Eq. (\ref{damptrans}), to the probability distribution we get $P_p=\langle N \rangle^{p}/(1+\langle N \rangle)^{p+1}$ and hence the effective temperature
\begin{equation}
T_\text{eff}=-\frac{\hbar \omega}{k_\text B} \ln \left[\frac{2\lambda^2\Gamma_\text L\Gamma_\text R}{\kappa(\Gamma_\text L+ \Gamma_\text R)+2\lambda^2\Gamma_\text L\Gamma_\text R}\right],
\label{Teff}
\end{equation}
valid for arbitrary $\lambda$. We point out that a tunneling induced thermal resonator state was also discussed in Refs. \onlinecite{Mitra-2004aa,Bergenfeldt-2012aa,Jin-2015aa}.

\section{Low frequency statistics}
\label{sec_longt}
%

\subsection{Currents and current correlations}
With the theory and method in place we turn to the analysis of the low frequency statistics. We first consider the currents and current correlations.

\subsubsection{Electronic cumulants}
For completeness, we start by considering the purely electronic cumulants $I_\text e$ and $S_\text{e,e}$. Due to the high bias limit, the total probability for electron tunneling  is the same as for an isolated quantum dot, i.e. independent of the resonator photon state. From  Eqs. (\ref{eq_IaExpansion}) and  (\ref{eq_SabExpansion}) we find, in agreement with the well known \cite{Bagrets-2003aa} results, the current
\begin{align}
\label{eq_Ie}
I_\text e=e\frac{\Gamma_\text L\Gamma_\text R}{\Gamma_\text L+\Gamma_\text R}
\end{align}
and the noise
\begin{align}
\label{eq_See}
S_\text{e,e}&=e^2 \Gamma_\text L \Gamma_\text R\frac{\Gamma_\text L^2+\Gamma_\text R^2}{(\Gamma_\text L+\Gamma_\text R)^3} \equiv e I_\text{e}F_\text{e},
\end{align}
where 
\begin{align}
F_\text{e}=\frac{\Gamma_\text L^2+\Gamma_\text R^2}{(\Gamma_\text L+\Gamma_\text R)^2}
\end{align}
is the electronic Fano factor. 

\subsubsection{Photonic cumulants}

For the purely photonic cumulants we get, from Eq. (\ref{eq_IaExpansion}), the average current
\begin{align}
I_\st p =  \frac{2\lambda^2}{e} I_\st e.
\label{photcurr}
\end{align}
This simple relation between $I_\text e$ and $I_\text p$ follows directly from the discussion of the average number of resonator photons: the rate for photon injection, equal to the photon current, is $2\lambda^2 I_\text e/e$. We stress that the relation between $I_\text e$ and $I_\text p$ in Eq. (\ref{photcurr}) is  valid for arbitrarily large $\lambda$ and independent of the resonator decay rate $\kappa$. Hence, Eq. (\ref{photcurr}) gives a direct and robust way to access the coupling strength $\lambda$ via average photon and electron current measurements.

\begin{figure}[t!]
\includegraphics[width=\linewidth]{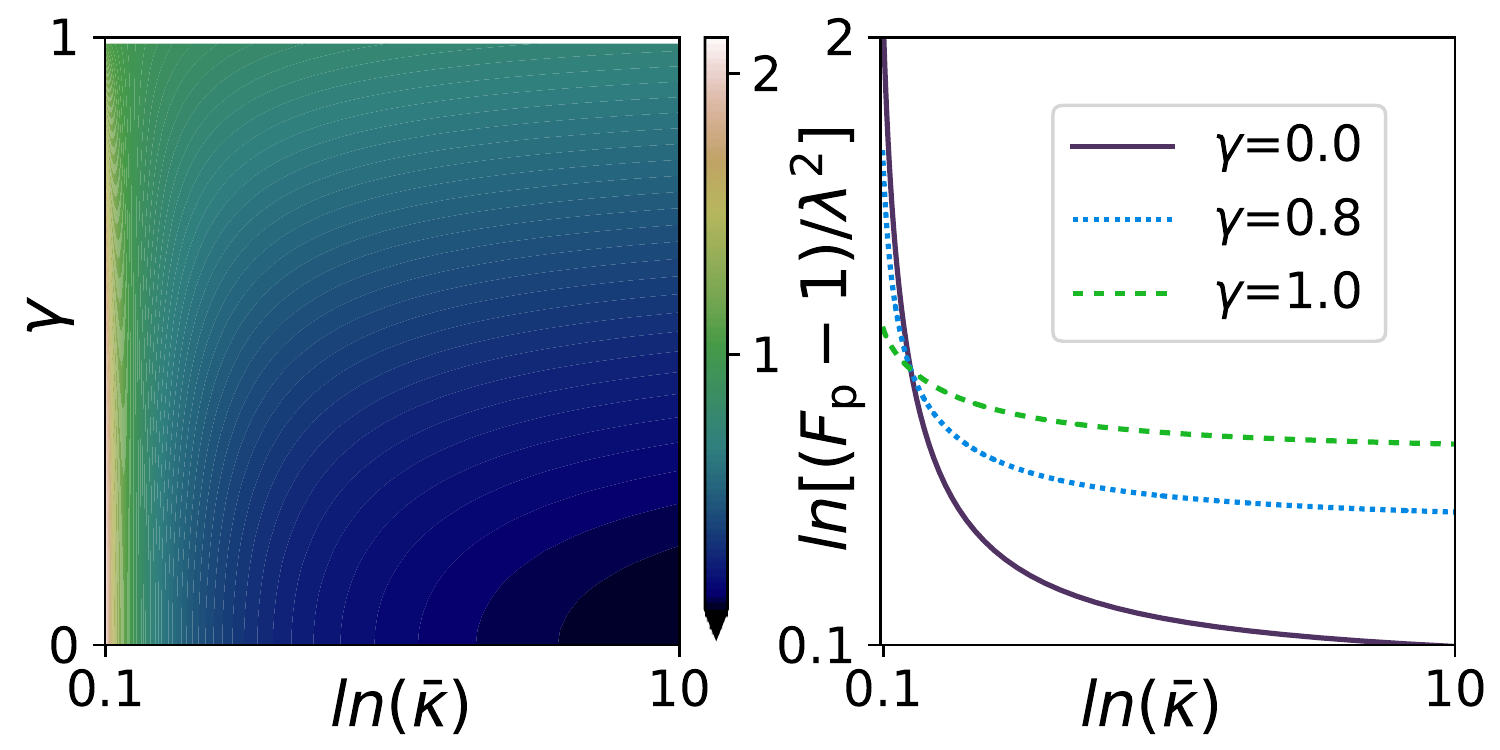}
 \caption{\label{fig_FanoSpp} Left panel: Map of \(\st{ln}[(F_\st p - 1)/\lambda^2]\), with $F_\st p - 1$ is the super-poissonian part of the photonic Fano factor \(F_\st p \), as a function of the dimensionless resonator decay rate \(\bar \kappa =\kappa/(\Gamma_\text L+\Gamma_\text R)\) and the tunneling asymmetry parameter  \( \gamma =( \Gamma_\text L -\Gamma_\text R)/(\Gamma_\text L+\Gamma_\text R)\). Right panel: Plots of \(\st{ln}[(F_\st p - 1)/\lambda^2]\) (same as left panel) as a function of $\bar \kappa$ for a number of representative $\gamma$.}
\label{photonfano}
\end{figure}

For the photon correlations, from Eq.~(\ref{eq_SabExpansion}) we get
\begin{align}
S_{\st p,\st p} =  I_\st p F_\st{p}
\label{photnoise}
\end{align}
with the photonic fano factor $F_\text p$ given by 
\begin{align}
F_\st p = 1+ \frac{\gamma^2 [\bar \kappa(\bar \kappa+1)-1]+(\bar \kappa+1)^2}{\bar \kappa (1+\bar \kappa)} \lambda^2.
\label{eq_Spp}
\end{align}
Here we introduced for convenience the dimensionless tunneling asymmetry parameter  $-1 \leq \gamma \leq 1$ and the renormalized photon decay rate $\bar \kappa$, defined as
\begin{align}
\gamma=\frac{\Gamma_\text L-\Gamma_\text R}{\Gamma_\text L+\Gamma_\text R}, \quad \bar \kappa=\frac{\kappa}{\Gamma_\text L+\Gamma_\text R}. 
\label{eq_redef}
\end{align}
Several interesting properties of $F_\text p$ can be noted from Eq. (\ref{eq_Spp}) and seen in the plot in Fig. (\ref{photonfano}). First, for all values of $\lambda, \kappa, \Gamma_\text L, \Gamma_\text R$ we have $F_\st p \geq 1$, i.e. the photon statistics is always super poissonian. Second, $F_\st p$ increases linearly with increasing $\lambda^2$ and is close to unity in the limit of small coupling parameter, $\lambda \ll 1$. Third, $F_\text p$ is a function of $\gamma^2$ only, i.e. symmetric in $\gamma$, and increases [decreases] linearly with $\gamma^2$ for $\bar \kappa > (\sqrt{5}-1)/2$ [$\bar \kappa <(\sqrt{5}-1)/2$]. Forth, as a function of the decay rate the Fano factor reaches $F_\text p=1+\lambda^2(1+\gamma^2)$ for $\bar \kappa \rightarrow \infty$. For $\bar \kappa \ll 1$, when the resonator photon state is thermal,  we have $F_\text p=1+\lambda^2(1-\gamma^2)/\bar \kappa=1+2\langle N \rangle$. For an interesting comparison see the recent ref.~\onlinecite{Brange_2019aa}.

\subsubsection{Electron-Photonic cross correlations}

Turning to the electron-photon cross correlations, the general expression from (\ref{eq_SabExpansion}), in terms of  $c_n^p$, $n,p=0,1$, is lengthy and we only give the resulting form, as
\begin{align}
S_{\st e,\st p} =& 2 e \Gamma_\text L \Gamma_\text R \frac{ \Gamma_\text L^2+\Gamma_\text R^2}{(\Gamma_\text L+\Gamma_\text R)^3} \lambda^2 = \frac{2\lambda^2}{e}S_\text{e,e}
\label{eq_Sep}
\end{align}
The similarity to Eq. (\ref{photcurr}) shows that the electron-photon cross correlations are governed by the pure electron noise, a consequence of that photons are added to the resonator, independent of the resonator state, with probability $\lambda^2$, for every electron tunnel event. Hence, beyond an additional way to determine $\lambda$ by comparing $S_\text{e,p}$ to $S_\text{e,e}$, there is no additional information about the electron-photon interaction in $S_\text{e,p}$ compared to $S_\text{e,e}$.

\begin{figure}
  \includegraphics[width=0.98\linewidth]{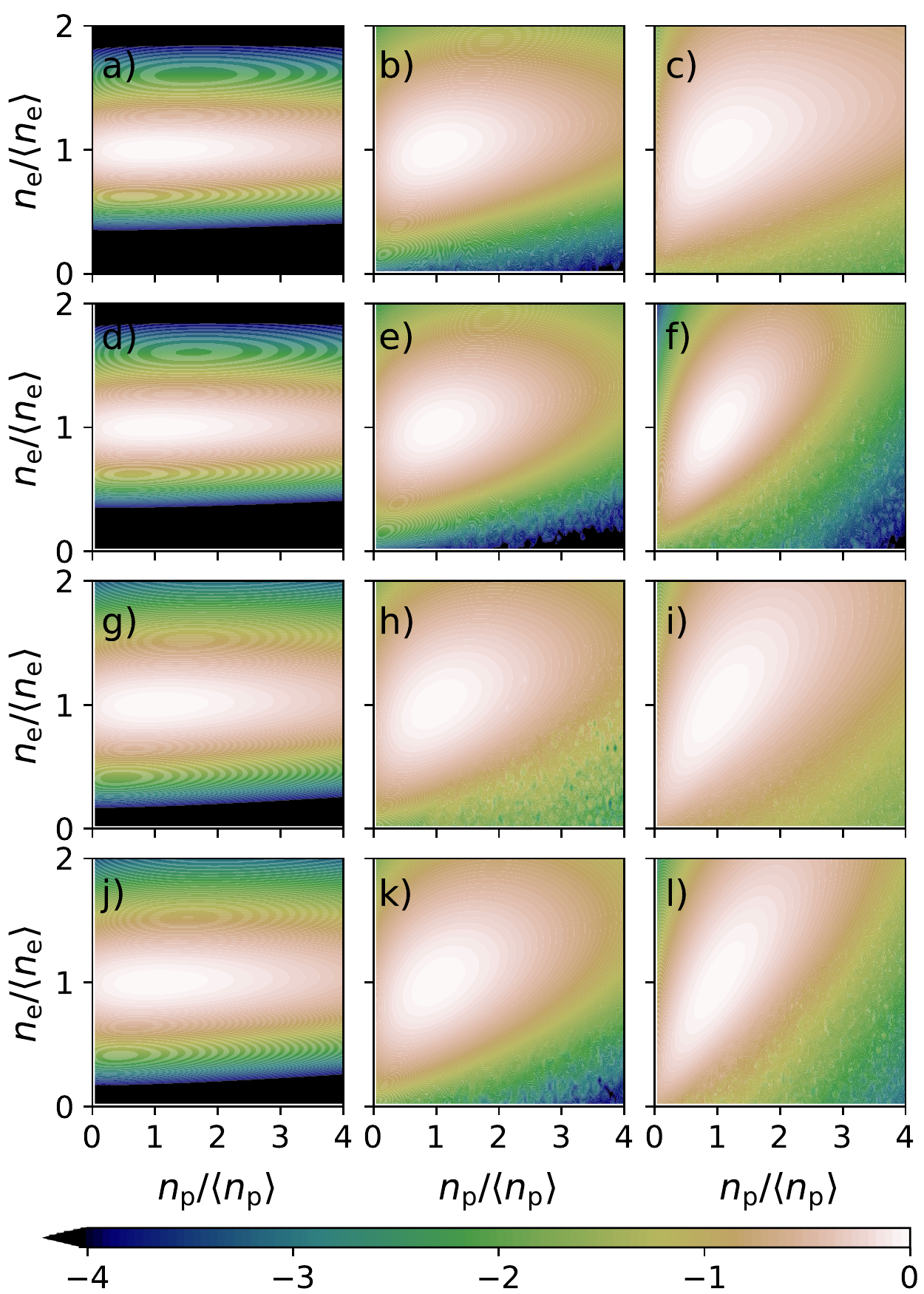}
  \caption{The joint electron-photon large deviation function \(\st{ln}[P(n_\st p , n_\st e )]/\sqrt{\langle n_\st p \rangle \langle n_\st e \rangle }\) as a function of $n_\text e/\langle n_\text e \rangle$ and  $n_\text p/\langle n_\text p \rangle$ for different sets of $\gamma, \bar \kappa$ and $\lambda$. In all rows the coupling parameter $\lambda$ increases from left to right, with $\lambda=0.1$ for a),d),g),j), $\lambda=0.4$ for b), e), h), k), and $\lambda=0.4$ for c), f), i), l). The assymmetry parameter is $\gamma=0$ for upper six panels, a)-f), and $\gamma=0.9$ for lower six panels, g)-l). The decay rate $\bar \kappa=0.5$  in the first and third row, a)-c) and g)-i), and $\bar \kappa=5$  in the second and fourth row, d)-f) and j)-l).}
\label{fig_lnP}
\end{figure}

\subsection{Full statistics}

The full, joint electron-photon probability distribution $P(n_\text e,n_\text p)$, plotted for a set of different parameters in Fig. \ref{fig_lnP} and further discussed below, is typically difficult to evaluate analytically. However, in certain limits we can get an explicit expression for the corresponding CGF \(F(\xi_\text e,\xi_\text p)\), which gives us additional insight into the particle transfer statistics. We discuss the different cases one by one below. 

\subsubsection{Weak coupling limit, $\lambda^2 \ll 1$}
For  $\lambda^2 \ll 1$, we may expand \(F(\xi_\text e,\,\xi_\text p)\) in order of \(\lambda^2\), as 
\begin{align}
\label{eq_Flambda}
F(\xi_\text e,\xi_\text p)=F_0(\xi_\text e)\! +\! \lambda^2F_1(\xi_\text e,\xi_\text p)\!+ \!\mathcal{O}(\lambda^4),
\end{align}
and solve the determinant equation \(\det [M(\xi_\text e,\xi_\text p) - F(\xi_\text e,\xi_\text p)]=0\) order by order. For the zeroth order, $\lambda=0$, we find
\begin{align}
\frac{F_0(\xi_\text e)}{\Gamma_\text L+\Gamma_\text R}= -\frac{1}{2} \left( 1- \sqrt{1+ \bar \gamma^2\left( \E^{\I \xi_\text e} -1\right)} \right) \
\label{F0}
\end{align}
which is just the known CGF for electron transport through a quantum dot \cite{Bagrets-2003aa}, not coupled to a resonator. Here we for shortness introduced $\bar \gamma^2=1-\gamma^2$. To first order in $\lambda^2$ we get
\begin{align}
\frac{F_1(\xi_\text e,\xi_\text p)}{\Gamma_\text L+\Gamma_\text R}=\frac 1 2 \frac{\bar \gamma ^2 \E^{\I \xi_\st e } (\E^{\I \xi_\st p }-1
)}{\sqrt{1+ \bar \gamma ^2 (\E^{\I \xi_\st e } - 1 )}}. 
\label{F1}
\end{align}
This describes the transfer of a single photon, via $e^{i\xi_\text p}$, due to the tunneling of the electrons. We note that neither $F_0(\xi_\text e)$ nor $F_1(\xi_\text e,\xi_\text p)$ depend on the loss rate $\kappa$. This $\kappa$-independence for $\lambda^2\ll 1$ is also seen in the plot of  $P(n_\text e,n_\text p)$ in Fig. \ref{fig_lnP}. However, going to higher orders, the terms proportional to $\lambda^4$ or higher in general depend on both $\Gamma_\text L,\Gamma_\text R$ and $\kappa$. These terms are increasingly long and difficult to interpret in a physically transparent way and are hence not presented here. 

\subsubsection{Large loss rate limit}
In the limit of large resonator leakage,  \(\bar \kappa = \kappa/(\Gamma_L+\Gamma_\text R)\gg 1\), the photons created at an electron tunneling will leave the resonator well before the next electron tunnel event occurs. For the tunneling electrons, the resonator will then be effectively empty at all times, and the steady state vector reduces to \(\bm P = (P^0_0,P^1_0)\), for an empty (\(P^0_0\)) or filled dot (\(P^1_0\)) and an empty resonator. Consequently, the only relevant Franck-Condon factors are \([M_\lambda]_{1q}=\E^{-\lambda^2/2} \lambda^{q-1}/\sqrt{(q-1)!}\) representing the transition probabilities between zero and \(q-1 \)  photons. Noting that the series \(\sum_{q=1}^{\infty} [M_\lambda]_{1q}^2  \E^{\I (q-1) \xi_\text p} = \E^{\lambda^2(\E^{\I \xi_\text p}-1)}\), the effective $2\times 2$ evolution matrix for the system can then be written 
\begin{align}
\label{eq_Mlargekappa}
M(\xi_\text e,\xi_\text p)= \begin{pmatrix} -\Gamma_\text L & \Gamma_\text R \E^{\lambda^2(\E^{\I \xi_\text p}-1) + \I \xi_\text e} \\ \Gamma_\text L \E^{\lambda^2(\E^{\I \xi_\text p}-1)} & -\Gamma_\text R \end{pmatrix}\,,
\end{align}
giving the CGF as
\begin{align}
\label{eq_CGF}
\frac{F(\xi_\text e,\xi_\text p)}{\Gamma_\text L+\Gamma_\text R}= -\frac{1}{2} \left( 1- \sqrt{1+ \bar \gamma^2\left(\E^{\I \xi_\text e +2 \lambda^2 (\E^{\I \xi_\text p}-1)} -1\right)} \right). \nonumber \\
\end{align}
Comparing this CGF to Eq. (\ref{F0}), the bare quantum dot electron transport one, it is clear that every electron tunneling through the dot gives rise to a cascade of photons, described by the multiplicative factor $ \E^{2 \lambda^2 (\E^{\I \xi_\text p}-1)}$. This factor formally constitutes the moment generating function of the number of photons emitted per cascade. The probability to have $k$ photons emitted in a cascade is thus $e^{-\lambda^2} \lambda^{2k}/k!$, i.e., a Poisson distribution with average number of emitted photons equal to $\lambda^2$. In the limit of large asymmetry, $\bar \gamma^2 \ll 1$, the joint electron-photon cascades events occur in an uncorrelated fashion. 

From the CGF in Eq. (\ref{eq_CGF}) and Eqs. (\ref{eq_saddlepoint}) and (\ref{eq_chisaddle}) we get the joint probability distribution, to exponential accuracy, as 

\begin{align}
\label{Plowk}
\frac{\st{ln} \ P (\tilde n _\st e , \tilde n_\st p )}{t(\Gamma_\st L + \Gamma_\st R )/2}
&= -1-\tilde n _e \frac{\bar \gamma ^2}{2} \st{ln} [H(\tilde n _\st e )] +
\sqrt{\gamma^2 + \bar \gamma^2 H[\tilde n _\st e ]} \nonumber \\
&+ \bar \gamma^2 \left( \tilde n _\st p \Big[1+\st{ln} \left( \frac{\tilde
n _\st e }{\tilde n _\st p }\right)\Big]-\tilde n _\st e \right)
\end{align}%
where \(\tilde n _a = n_a/\langle n_a \rangle \), and the average number of electrons and photons are $\langle n_\text e \rangle=tI_\text e/e=t(\Gamma_\text L+\Gamma_\text R)\bar \gamma^2/4$ and $\langle n_\text p \rangle=tI_\text p=t(\Gamma_\text L+\Gamma_\text R)\bar \gamma^2\lambda^2/2$ and we introduced  \(H(x)= x(x \bar \gamma ^2 /2 + \sqrt{ \gamma^2 + (x \bar \gamma ^2 /2)^2})\). We note that already for $\bar \kappa=5$, as shown in Fig. \ref{fig_lnP}, the full numerical result and the limiting expression in Eq. (\ref{Plowk}) (not plotted here) are very similar, showing that the large damping limit is effectively reached already for moderately large $\bar \kappa$. 

\subsubsection{Marginal distributions}
The marginal electron and photon probability distributions are defined as $P_\text e(n_\text e)=\sum_{n_\text p}  P(n_\text e,n_\text p)$ and $P_\text p(n_\text p)=\sum_{n_\text e}  P(n_\text e,n_\text p)$, respectively. Their corresponding CGFs are given from the full $F(\xi_\text e,\xi_\text p)$ as $F_\text e(\xi_\text e)=F(\xi_\text e,0)$ and $F_\text p(\xi_\text p)=F(0,\xi_\text p)$. For the electron distribution, one can sum over the photon degree of freedom in Eq. (\ref{eq_MP=FP}). Noting that $\sum_k [M_\lambda]_{kq}=1$ and writing  $P^0(\xi_\text e)=\sum_p P^0_p(\xi_\text e,0)$ and  $P^1(\xi_\text e)=\sum_p P^1_p(\xi_\text e,0)$, we get
\begin{align}
\begin{pmatrix} -\Gamma_L & \Gamma_\text R \E^{i\xi_\text e} \\ \Gamma_L & -\Gamma_\text R \end{pmatrix} \begin{pmatrix} P^0(\xi_\text e) \\  P^1(\xi_\text e) \end{pmatrix} = F_\text e (\xi_\text e) \begin{pmatrix} P^0(\xi_\text e) \\  P^1(\xi_\text e) \end{pmatrix}
\end{align}  
and hence the  CGF
\begin{align}
F_\text e (\xi_\text e)= F_0  (\xi_\text e)
\end{align}
where $F_0  (\xi_\text e)$ is given by Eq. (\ref{F0}), i.e., the result for a quantum dot, not coupled to a resonator. This is in line with the findings for the electron current and noise above.

\begin{figure}
  \includegraphics[width=0.98\linewidth]{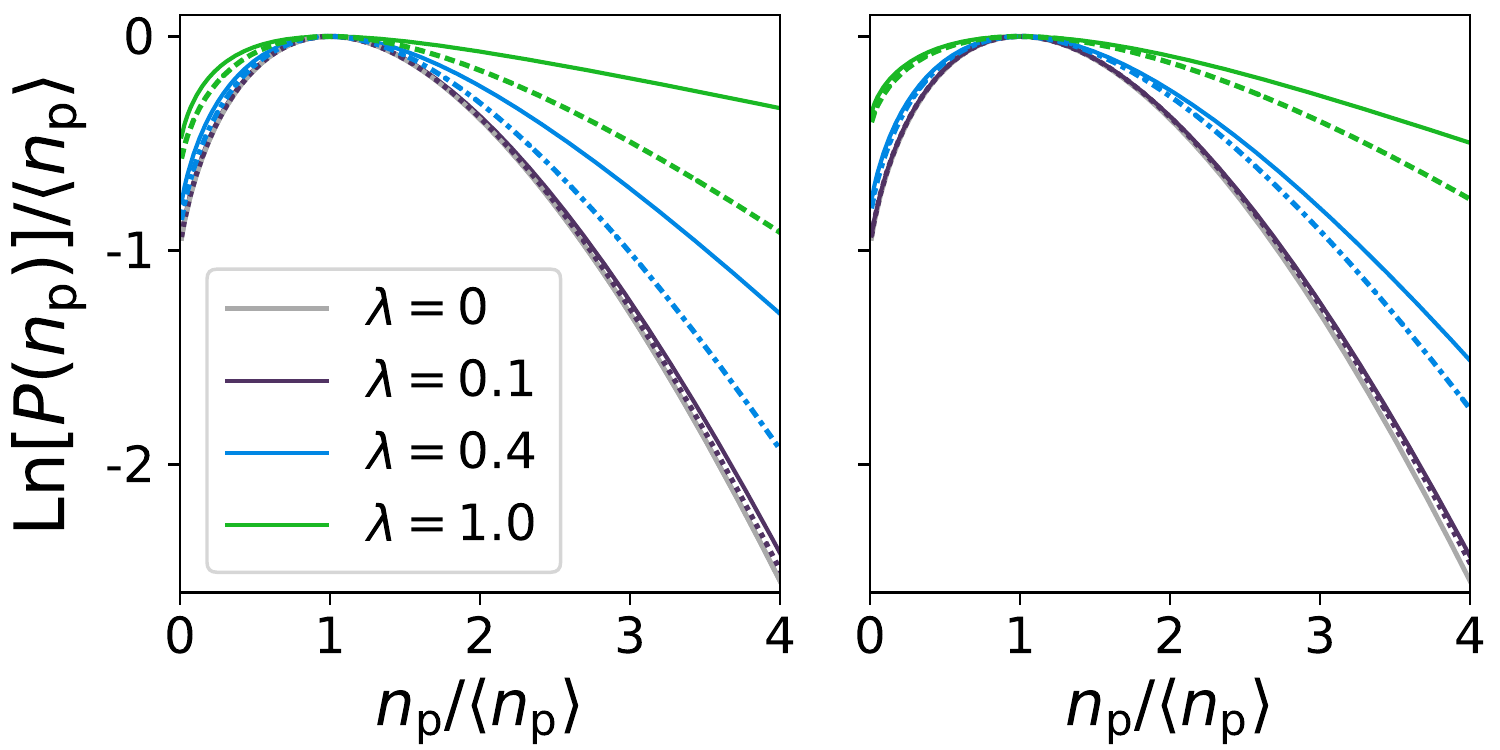}
\caption{\label{fig_lnPphoton} The marginal photon large deviation function \(\st{ln}[P_\text p(n_\st p)]/\langle n_\st p \rangle\) as a function of $n_\text p/\langle n_\text p \rangle$, for a set of couplings $\lambda$ and resonator decay rate \(\bar \kappa=0.5\)  (solid lines) and $\bar \kappa=5$ (dashed lines). The dot asymmetry parameter is $\gamma=0$ (left panel) and $\gamma=0.99$ (right panel).}
\end{figure}

The photon distribution is plotted for a representative set of parameters in Fig. \ref{fig_lnPphoton}. In the limit $\lambda^2 \ll 1$ the distribution becomes independent of $\bar \kappa$, as noted above. To first order in $\lambda^2$ we have from Eqs. (\ref{F0}) and (\ref{F0})
\begin{align}
\frac{F_\text p (\xi_\text p)}{\Gamma_\text L+\Gamma_\text R}=\frac{\lambda^2\bar \gamma^2}{2}\left(e^{i\xi_\text p}-1\right)
\end{align}
that is, a Poissonian distribution of photons, with emission rate $\lambda^2\bar \gamma^2/2$, consistent with the results for the photon current $I_\text p$, Eq. (\ref{photcurr}), and noise $S_\text p$, Eq. (\ref{photnoise}). The probability distribution $P(n_\text p)$, is thus given by 
\begin{align}
\frac{\ln P_\text p(\tilde n _\text p)}{\langle n_\text p\rangle}=-\tilde n _\text p \ln \left[\tilde n _\text p \right]+\left(\tilde n _\text p-1\right)
\end{align}
as seen in Fig. \ref{fig_lnPphoton}. We note that higher order terms $\propto \lambda^{2q}$, $q \geq 2$ in the CGF are proportional to $(e^{i\xi_\text p}-1)^q$, showing that multi-photon physics in the long-time limit is relevant only away from the weak coupling limit.

\subsubsection{Full results, numerical evaluation}
Outside the limiting regimes described above the joint electron-photon distribution function \(P(n_\st p , n_\st e)\) is evaluated numerically. The result is plotted in Fig.~\ref{fig_lnP}, for a representative set of $\lambda, \bar \kappa, \gamma$. We see that the probability distribution interpolates smoothly between the limiting cases discussed above, peaking aroung $n_\st p \sim \langle n_\st p \rangle, n_\st e \sim \langle n_\st e\rangle$ with a width depending on the system parameters.

We point out that in experiments, microwave photon detector efficiencies are typically non-unity,  $\eta < 1$. To account for this, in all expressions above one can replace 
\begin{equation}
e^{i\xi_\text p}\rightarrow 1-\eta +\eta e^{i\xi_\text p}
\end{equation}
This describes a binomial statistics of successful detection events for the emitted photons, with probability $\eta$ to detect a photon and $1-\eta$ to miss it.

\section{Short time correlations}
\label{sec_g2}
To evaluate the short time correlators we first write Eq.~(\ref{eq_g2definition}) in the damping basis, giving
\begin{align}
\label{eq_dampingbasisg2pp}
g^{(2)}_{\st a,\st b} (t) = \frac{ \bar{\bm v}_0 \bar M'_\text a \E^{-\bar M t}  \bar M'_\text b \bar{\bm P}}{I_\st a I_\st b}
\end{align}
where we write $\bar M=\bar M(0,0)$. The lower-triangular-per-block structure of the matrices $\bar M'_\text a, \bar M$, in the damping basis, and the specific structure of $\bar{\bm P}$ are described in Eqs. (\ref{damptrans}) to (\ref{eq_barP}). These properties allow us to both evaluate the $g^{(2)}$-functions analytically, for arbitrary $\lambda$, and to provide a physically transparent picture of the result. Most importantly, the $\lambda$-dependence of the numerator of the correlator is directly determined by the matrices $\bar M ' _\st a$; a matrix $\bar M ' _\st e$ , describing an electronic tunneling event, contributes with a $\lambda$-indepent factor while the photonic emission matrix $\bar M ' _\st h$ contributes with a factor $\propto \lambda^2$. As a result the numerator in Eq. (\ref{eq_dampingbasisg2pp}) is proportional to $1, \lambda^2$ and $\lambda^4$ for $\{\st{a,b}\}=\{\st{e,e}\}, \{\st{e/p,p/e}\}$ and $\{\st{p,p}\}$ respectively. However, the currents in the denominator, given in Eqs. (\ref{eq_Ie}) and (\ref{photcurr}), are independent of $\lambda$ (for $I_\st e$) and proprtional to $\lambda^2$ (for $I_\st p$). As a consequence, all  $g^{(2)}$-functions, electronic, photonic as well as the cross correlations, are independent of $\lambda$. This is in stark contrast to the long time correlators $S_\st{a,b}$ discussed above, clearly showing the difference in physical information contained in the short and long time correlations. Below we discuss the different correlators separately, introducing for convenience the dimensionless time \(\tilde t = (\Gamma_\text L+\Gamma_\text R) t\).

\subsubsection{Electron-electron correlations}
For completeness we give the electronic function,  becoming
\begin{align}
\label{eq_dampingbasisg2pp}
g^{(2)}_{\st e,\st e} (t) =1-e^{-\tilde t}
\end{align}
which is the result found \cite{Emary-2012aa} for a quantum dot, not coupled to a resonator.  Similar to the findings above, in the high bias limit considered, the resonator has no effect on the bare electronic transport (leaving the photons unobserved). We note that $g^{(2)}_{\st e,\st e} (0)\leq g^{(2)}_{\st e,\st e} (t)\leq 1$, describes the anti-bunching of electrons tunneling through the dot, a consequence of that an electron in the dot has to tunnel out before another electron can enter.

\subsubsection{Photon-photon correlations}
For the photon correlation function we get 
that
\begin{align}
\label{eq_g2pp}
g^{(2)}_{\st p,\st p} (t ) &= 1-\frac{\gamma^2\bar \kappa^2}{(1-\gamma^2)(1-\bar\kappa^2)}\E^{-\tilde t } \nonumber \\
 &  +\frac{(1-\gamma^2)+\bar\kappa\left[(1+2 \gamma^2)-\bar \kappa-\bar \kappa^2\right]}{(1-\gamma^2)(1-\bar\kappa^2)}\E^{-\bar\kappa \tilde t }
\end{align}
This expression shows that the dynamics of the photon emission is governed by two rates, $\Gamma_\text L+ \Gamma_\text R$ and $\kappa$. Interestingly, for a symmetric dot, $\gamma=0$, we have $g^{(2)}_{\st p,\st p} (t )=1+(1+\bar \kappa)e^{-\bar \kappa \tilde t}$, with the dynamics goverened only by the resonator decay rate $\kappa$. By rewriting Eq. (\ref{eq_g2pp}) as $[g^{(2)}_{\st p,\st p} (t )-1]\bar \gamma^2(1+\bar \kappa)=\gamma^2\bar \kappa^2(e^{-\bar \kappa \tilde t}-e^{-\tilde t})/(1-\bar \kappa)+[\bar \gamma^2+\bar \kappa(2+\gamma^2+\bar \kappa)]e^{-\tilde t}$, we note the following. First,  $g^{(2)}_{\st p,\st p} (t )-1\geq 0$, i.e., $g^{(2)}_{\st p,\st p} (t )\geq 1$. Second, $g^{(2)}_{\st p,\st p} (t)$ is monotonically decreasing with $t$, for any $\gamma,\bar \kappa$. Third, as is not directly apparent from Eq. (\ref{eq_g2pp}),  $g^{(2)}_{\st p,\st p} (t)$ is well behaved for $\bar \kappa=1$. 

Based on these observations we can conclude that the emitted photons, in contrast to the emitted electrons,  are bunched, $g^{(2)}_{\st p,\st p} (0) \geq g^{(2)}_{\st p,\st p} (t) \geq 1$, and hence that the emitted radiation is classical, for all system parameters $\lambda, \gamma,\kappa$. Moreover, due to the monotonic-in-time behavior it is mainly interesting to analyze the $t=0$ correlator, given by
\begin{align}
\label{eq_g2pptzero}
g^{(2)}_{\st p,\st p} (0) &=2+\frac{2(1-\bar \gamma^2)+(1+\bar \kappa)}{\bar \gamma^2(1+\bar \kappa)}\bar \kappa
\end{align}
We see that for $\bar \kappa \gg \bar \gamma^2$, the correlator becomes large, $g^{(2)}_{\st p,\st p} (0) \approx \bar \kappa/\bar \gamma^2 \gg 1$. This limit can be reached for a strongly asymmeric dot,  $\bar \gamma^2 \ll 1$ and/or large effective resonator decay $\bar \kappa \gg 1$. The physical mechanism behind the large correlator in this parameter limit is the uncorrelated cascades of photons emitted into a leaky resonator, which gives a large probability to observe more than one photon emitted from the resonator at the same time. We stress that Eq. (\ref{eq_g2pptzero}) holds for any $\lambda$, hence, the large $g^{(2)}_{\st p,\st p} (0)$ is expected also  in the weak coupling limit $\lambda^2\ll 1$, where photon Fano factor, Eq. (\ref{photonfano}), approaches unity.    

\begin{figure}
\includegraphics[width=\linewidth]{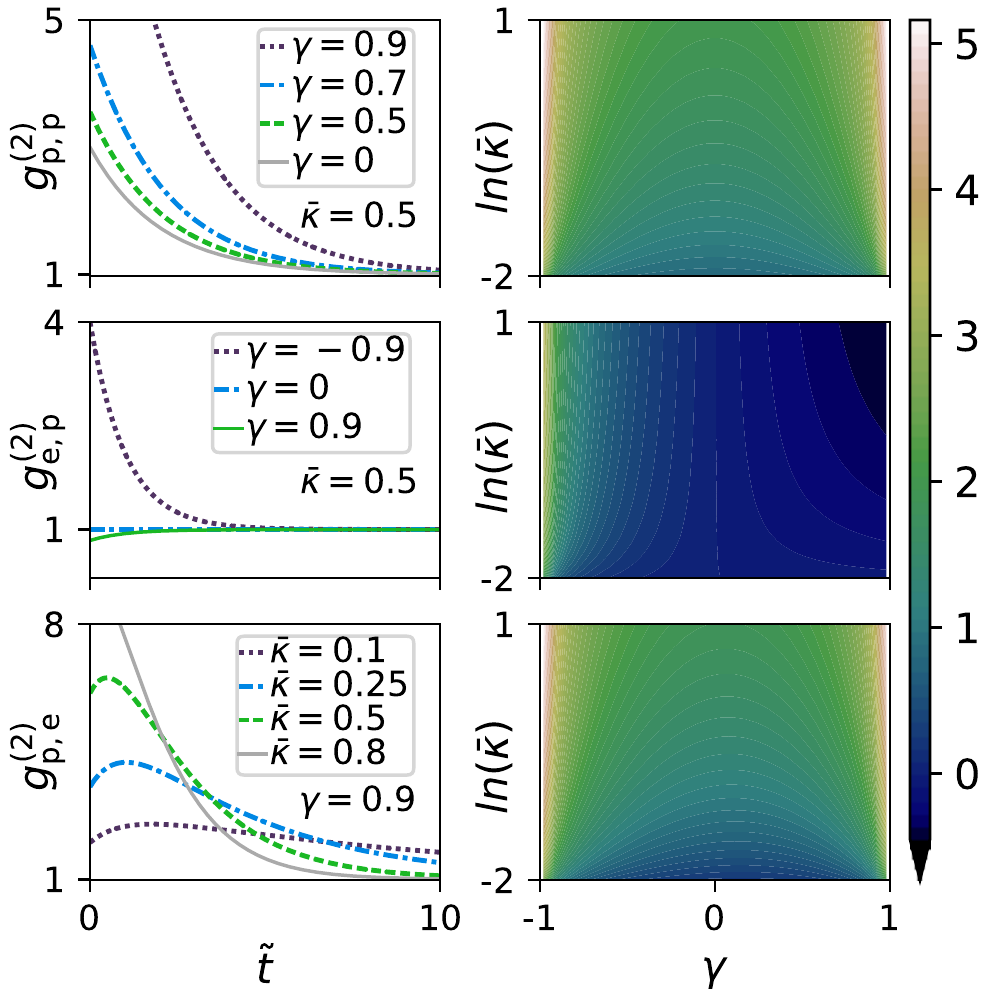}
\caption{Left panels: Short time correlations functions $g^{(2)}_{\st p, \st p} ( t )$ (top), $g^{(2)}_{\st e, \st p} ( t )$ (middle) and $g^{(2)}_{\st p, \st e} ( t )$ (bottom) as a function of dimensionless time $\tilde t=t/(\Gamma_\text L+\Gamma_\text R)$, for sets of representative $\bar \kappa$ and $\gamma$. Right panels: Maps of zero time correlations functions $\ln[g^{(2)}_{\st p, \st p} (0)]$ (top), $\ln[g^{(2)}_{\st e, \st p} (0)]$ (middle), and $\ln[g^{(2)}_{\st p, \st e} (0)]$ (bottom),  as a function of $\bar \kappa$ and $\gamma$.}
\label{fig_g2}
\end{figure}

From Eq. (\ref{eq_g2pptzero}) it is clear that for a small resonator decay rate, $\bar \kappa \ll 1$, when the resonator photon state is thermal, we have $g^{(2)}_{\st p,\st p} (0)=2$. In fact, the full time dependence of the correlator in this limit is $g^{(2)}_{\st p,\st p} (t )=1+e^{-\kappa t}$, which is the known\cite{Walls} result for emission out from a resonator in a thermal state. To illustrate the results in Eq. (\ref{eq_g2pp}) and (\ref{eq_g2pptzero}), in Fig. \ref{fig_g2} we plot both the time dependence of the correlator as well as the $t=0$ result, for different representative parameters.

\subsubsection{Cross correlations}
Turning to the electron-photon cross correlations, we recall that we consider the electronic measurement at the left lead (L), i.e., when the electron enters the dot. If we instead would measure electrons leaving the dot, all results below would be modified by taking \(\gamma \rightarrow - \gamma\). We find 
\begin{align}
g^{(2)}_{\st e, \st p} ( t ) 
= 1&-\frac{\gamma \bar \kappa}{(1+\gamma)(1+\bar \kappa)}\E^{-\tilde t} \\
g^{(2)}_{\st p, \st e} (t ) 
= 1 & -\frac{\gamma \bar \kappa}{(1-\gamma)(1-\bar \kappa)}\E^{-\tilde t} \nonumber \\
&+ 2 \bar \kappa\frac{1+\gamma^2+(\gamma-\bar \kappa)\bar\kappa}{(1-\gamma^2)(1-\bar \kappa ^2)}\E^{-\bar \kappa \tilde t }\label{eq_g2pe}
\end{align}
We stress that the two correlators are not equal to each other, $g^{(2)}_{\st e, \st p} (t ) \neq g^{(2)}_{\st p, \st e} (t )$. It hence makes a difference which particle is detected first, the electron or the photon, and we therefore discuss the two cases separately. 

For $g^{(2)}_{\st e, \st p} ( t )$ one first measures a photon at $t=0$, and thereafter waits for an electron to enter the dot from the left lead. The correlator is bounded from below, $1/2 \leq g^{(2)}_{\st e, \st p} ( t)$, and decays (increases) monotonically with time for an asymmetry $\gamma <0$ ($\gamma >0$), with the rate $\Gamma_\text L+ \Gamma_\text R$. For $t=0$ we have $g^{(2)}_{\st e, \st p} (0)=1-\gamma \bar \kappa/[(1+\gamma)(1+\bar \kappa)]$. The shift from an anti-bunching-like behavior, for $\gamma >0$, to a bunching-like behavior, for $\gamma < 0$, reflects the underlying correlations between electrons tunneling at contacts L and R. In particular, for large asymmetry $\gamma \rightarrow -1$, the dot is empty most of the time and an in-tunneling event (at contact L) is rapidly followed by an out-tunneling event (at contact R), at which we measure. We then get \(g^{(2)}_{\st e, \st p} (0) \sim 1/(1+\gamma) \gg 1\).
 
For $g^{(2)}_{\st p, \st e} ( t )$ one instead first measures an electron passing from the right lead into the dot and thereafter waits for a photon to leave the resonator. The dynamics is governed by the rates $\Gamma_\text L+ \Gamma_\text R$ and $\kappa$, with the correlator dispalying a non-monotonic dependence on time. However, by rewriting the correlator similarly to $g^{(2)}_{\st p, \st p} ( t )$ above, one can show that $g^{(2)}_{\st p, \st e} ( t ) \geq 1$ for all times, thus showing a bunching-like behavior despite the fact that $g^{(2)}_{\st p, \st e} ( t ) \ngeq g^{(2)}_{\st p, \st e} (0)$ for some $\gamma, \bar\kappa$. At time $t=0$ we have  $g^{(2)}_{\st e, \st p} (0)=(1-\gamma+\bar \kappa)(1+\gamma+2\bar \kappa)/[(1-\gamma^2)(1+\bar\kappa)]$. Hence, in contrast to $g^{(2)}_{\st e, \st p} (0)$, the correlator becomes very large, $g^{(2)}_{\st p, \st e}(0) \sim 1/(1-\gamma^2)\gg 1$, in both asymmetric limits  $\gamma \rightarrow  \pm 1$. Moreover, independent of asymmetry $\gamma$, in the leaky resonator limit $\bar \kappa \gg 1$ the correlator also becomes large, $g^{(2)}_{\st p, \st e}(0) \sim \kappa$. A detailed investigation in the different mechanisms behind these large correlators goes beyond the present work.

As a consistency check, we note that integrating the \(g^{(2)}\) functions and using equation (\ref{eq_Sabisintg2}) we indeed find back the noise expressions as found in (\ref{eq_See}), (\ref{eq_Spp})-(\ref{eq_Sep}). 

\section{Comparison to related work}
\label{sec_compare}
Comparing to results in similar systems, we can make a number of observations. A thermal photon state in the resonator was found for a metallic dot \cite{Bergenfeldt-2012aa} and for a tunnel barrier \cite{Jin-2015aa} coupled to a resonator.  However, in refs. \onlinecite{Bergenfeldt-2012aa,Jin-2015aa}, the thermal state were obtained in the limit $\lambda \ll 1$, our result in Eq. (\ref{Teff}) holds for arbitrary $\lambda$. We also note that tunneling induced thermal states have been discussed for conductor-harmonic oscillator systems, see e.g. \cite{Mozyrsky-2002aa,Mitra-2004aa} for early works. 

For the transport properties we note that the high-bias limit makes the electron transport properties,  not detecting the emitted photons, insensitive to the presence of the resonator. As a consequence, the electron transport properties are the same as the ones for a quantum dot without the cavity, both the long \cite{Bagrets-2003aa} and short \cite{Emary-2012aa} time statistics. For the photon emission, not detecting the tunneling electrons, we note that the photons are bunched, with  super-poissonian statistics, for all system parameters. Similarly, bunching of emitted photons was predicted for a tunnel junction coupled to a tunnel junction \cite{Jin-2015aa}, for arbitrary $\lambda$. However, photon anti-bunching and sub-Poissonian statistics were predicted over a broad range of parameters for a resonator driven by a coherent double quantum dot \cite{Xu-2013aa}. A qualitative understanding of this difference is presently not clear.

\section{Conclusion and outlook}
\label{sec_conclusion}
We have investigated combined charge-photon statistics in a quantum dot-resonator system in the high bias and sequential tunneling limit. The focus has been on systems with large, dimensionless coupling parameter $\lambda \gtrsim 0.1$, recently realized in various experiments. Analyzing both the long-time and short-time statistics, we identify transport signatures of strong electron-photon coupling. This includes electron tunneling indiced cascaded photon emission, manifested in both the long time photon counting statistics as well as in short time $g^{(2)}$-functions. The coupling parameter $\lambda$ can be directly obtained from independent measurements of the average electrical and photon currents, while all electron and photon  $g^{(2)}$-functions are found to be independent of $\lambda$. We also find that for a small photon decay rate the resonator photon state becomes thermal, a property clearly manifested in the photon emission properties.

We hope our work will stimulate further theoretical investigations, in particularly focusing on the regime of large conductor-resonator coupling. One open problem, not clarified by our work, is to establish physically transparent conditions for when emitted resonator photons inherit the statistical properties of the tunneling electrons, that is, show anti-bunching and sub-Poissonian statistics. Moreover, the proposed system is one of the most elementary conductor-resonator ones. The calculated transport quantities are all experimentally measurable with present day technics or within reach. In addition, our results are mostly analytical and valid for arbitrarily large coupling parameter $\lambda$. Taken together, this makes our proposal highly relevant for experimental transport investigations of strong electron-photon interactions at the nanoscale. \\

\section{Acknowledgements}
We acknowledge discussions with J. Basset and K. Ensslin at an early stage of this project.  The  work  of  TB  is  supported  by  the  Spanish  Ministeriode Economía y Competitividad (MINECO) through the project  FIS2014-55987-P,  by  the  Spanish  Ministerio de Ciencia, Innovation y Universidades (MICINN) under the project FIS2017-82804-P. Both  TB and PS were supported by the Swedish Research Council, VR. 

\appendix

\section{Damping basis coefficients}

 The values of the lowest order coefficients \(c^n_p\) in Eq. (\ref{eq_barP}), needed for explicit evaluations below, are given by
\begin{eqnarray}
c^0_0&=&\frac{\Gamma_\text R}{\Gamma_\text L+\Gamma_\text R}, \quad c^1_0=\frac{\Gamma_\text L}{\Gamma_\text L+\Gamma_\text R} \label{eq_c00}\\ \nonumber
c^0_1&=&c_0^0\frac{\Gamma_\text L(2\Gamma_\text R+\kappa)}{\kappa (\Gamma_\text L+\Gamma_\text R+\kappa)}, \quad c^1_1=c^1_0\frac{\Gamma_\text R (2\Gamma_\text L+\kappa)}{\kappa (\Gamma_\text L+\Gamma_\text R+\kappa)} \\ \nonumber
c^0_2&=& c^0_1 \frac{\kappa(2 \Gamma_\text R^2+4\Gamma_\text R \kappa+\kappa^2) +\Gamma_\text L(4\Gamma_\text R^2+6\Gamma_\text R \kappa + \kappa^2) }{\kappa (2\Gamma_\text R+\kappa)(\Gamma_\text L+\Gamma_\text R+2\kappa)} \\  \nonumber
c^1_2&=&c^1_1 \frac{\kappa(2 \Gamma_\text L^2+4\Gamma_\text L \kappa+\kappa^2)+\Gamma_\text R(4\Gamma_\text L^2+6\Gamma_\text L \kappa + \kappa^2) }{\kappa (2\Gamma_\text L+\kappa)(\Gamma_\text L+\Gamma_\text R+2\kappa)}
\label{eq_c12}
\end{eqnarray}
Higher order coefficients can be found from lower ones by means of a recursion relation, here not further discussed. 

To illustrate how the transport properties are calculated, we present the expressions for the average currents and current correlations, Eqs (\ref{eq_IaExpansion}) and (\ref{eq_SabExpansion}), in terms of the coefficients  \(c^n_p\) . For the average currents we find
\begin{align}
I_\text e=e\Gamma_\text Rc_0^1=e\frac{\Gamma_\text L\Gamma_\text R}{\Gamma_\text L+\Gamma_\text R},
\end{align}
giving Eq. (\ref{eq_Ie}) in the main text, and 
\begin{align}
I_\st p = \kappa \lambda^2(c^0_1+c^1_1) =  \frac{2\lambda^2}{e} I_\st e.
\end{align}
giving Eq. (\ref{photcurr}). For the current correlators we have 
\begin{align}
S_\text{e,e}&= e^2 \frac{\Gamma_\text R}{\Gamma_\text L+\Gamma_\text R} c^1_0[\Gamma_\text L+\Gamma_\text R(c^0_0-c^1_0)]\\
&=e^2 \Gamma_\text L \Gamma_\text R\frac{\Gamma_\text L^2+\Gamma_\text R^2}{(\Gamma_\text L+\Gamma_\text R)^3},
\end{align}
giving Eq. (\ref{eq_See}) and
\begin{align}
S_{\st p,\st p} = & \kappa \lambda^2 \Bigg( (c^0_1+c^1_1) + \lambda^2\Bigg[ 4(c^0_2+c^1_2) -2(c^0_1+c^1_1)^2 \nonumber\\
+&\frac{\Gamma_\text L-\Gamma_\text R}{\Gamma_\text L+\Gamma_\text R}\left[(c^0_1-c^1_1)-(c^0_1+c^1_1)(c^0_0-c^1_0)\right]\Bigg] \Bigg) \nonumber\\
\equiv & I_\st p\left( 1+ \frac{\gamma^2 [\bar \kappa(\bar \kappa+1)-1]+(\bar \kappa+1)^2}{\bar \kappa (1+\bar \kappa)} \lambda^2\right)
\end{align}
giving Eq. (\ref{photnoise}). The electron-photon cross correlations can be expressed in a similar, but lengthy, way not presented here.

\bibliography{eletron_photon}

\end{document}